\documentstyle[epsf]{mn}
\begin{document}
\newcommand{\twiddles}{\sim}
\newcommand{\etal}{{et al}\/.}
\newcommand{\uv}{{\it uv}}
\def\Ssin#1C#2 {#1C\,#2}
\def\Ss#1{\Ssin#1 }
\def\Ssf#1{\Ssin#1 }
\def\afterpage#1{}
\def\dgr{^\circ}
\title[The WAT source 3C\,130]{Jets, plumes and hot spots in the wide-angle tail source \Ss{3C130}}
\author[M.J.~Hardcastle]{M.J.~Hardcastle$^1$\thanks{E-mail: M.Hardcastle@bristol.ac.uk}$^{,2}$\\
$^1$ H.H.\ Wills Physics Laboratory, University of Bristol, Royal Fort,
Tyndall Avenue, Bristol BS8 1TL\\
$^2$ Mullard Radio Astronomy Observatory, Cavendish
Laboratory, Madingley Road, Cambridge, CB3 0HE}
\maketitle
\begin{abstract}
I present 1.5- and 8.4-GHz observations with all configurations of the
NRAO VLA of the wide-angle tail source \Ss{3C130}. The source has a
pair of relatively symmetrical, well-collimated inner jets, one of
which terminates in a compact hot spot. Archival {\it ROSAT} PSPC data
confirm that 3C\,130's environment is a luminous cluster with little
sign of sub-structure in the X-ray-emitting plasma. I compare the
source to other wide-angle tail objects and discuss the properties of
the class as a whole. None of the currently popular models is entirely
satisfactory in accounting for the disruption of the jets in 3C\,130.
\end{abstract}
 
\section{Introduction}

\Ss{3C130} is a FRI radio source at redshift 0.109 (Spinrad \etal\
1985). Its 178-MHz luminosity is $7.6 \times 10^{25}$ W Hz$^{-1}$
sr$^{-1}$, slightly above the nominal FRI-FRII boundary of $\sim 2
\times 10^{25}$ W Hz$^{-1}$ sr$^{-1}$ (Fanaroff \& Riley 1974,
hereafter FR). Leahy (1985, 1993) and J\"agers and de Grijp (1985) present
intermediate-resolution VLA maps of the central regions of the source,
while J\"agers (1983) has a lower-resolution WSRT image which shows
the whole source and its field; the source extends for $\sim 1.5$
Mpc. Saripalli \etal\ (1996) present high-frequency maps made with the
Effelsberg 100-m telescope. The host galaxy is classed as a DE2 by
Wyndham (1966) and appears to lie in a cluster, although strong
galactic reddening makes optical identification of the cluster members
difficult. The {\it Einstein} detection of extended X-ray emission
(Miley \etal\ 1983), the near\-by align\-ed sources (J\"agers 1983)
and the many mJy radio sources in the field at 1.5 GHz make it
plausible that the object is the dominant member of a large
cluster. Leahy (1985) also attempts to constrain the RM distribution
of the source, but notes that it depolarizes rapidly (particularly in
the S lobe) so that few good measurements are available; this could be
taken as evidence for a dense magneto-ionic environment for the source
(cf.\ Hydra A, Taylor \etal\ 1990).

\label{definition}
3C\,130 is a wide-angle tail (WAT) radio source. The term WAT has been
used to describe many different types of object. Here I shall use it
to refer to those FRI sources which are associated with central
cluster galaxies (e.g.\ Owen \& Rudnick 1976) and have luminosities
comparable to or exceeding the Fanaroff-Riley break between FRI and
FRII. I shall follow Leahy (1993) in using the behaviour of the jets
at the base as another defining feature. At high resolution one or two
well-collimated jets [`strong-flavour' jets, by the classification of
Leahy (1993)] are seen (e.g.\ O'Donoghue, Owen \& Eilek 1990),
extending for some tens of kpc before broadening, often at a bright
flare point, into the characteristic plumes or tails. These jets are
very similar to the jets seen in FRII radio galaxies, and quite
different from the behaviour of jets in more typical FRIs, where a
collimated inner jet, if visible at all, decollimates rapidly (on
scales of a few kpc at most) and comparatively smoothly into a bright
`weak-flavour' jet with a large opening angle.\footnote{There are a
few exceptions to this behaviour; \Ss{3C66B} (Hardcastle \etal\ 1996)
does appear to show an inner `strong-flavour' jet and a bright knot at
the base of the `weak-flavour' jet. But even here the transition from
strong to weak flavours occurs on scales of $\sim 1$ kpc.} WATs,
according to this definition, never have a weak-flavour jet, but make
the transition between strong-flavour jet and diffuse, bent tail in a
single step. The requirement that WATs be central cluster galaxies
excludes objects (e.g. \Ss{3C171}, Blundell 1996, Hardcastle \etal\
1997a; \Ss{3C305}, Leahy 1997) where the `tails' are likely to be
simply ordinary FRII lobes which have been disrupted by unusual
host-galactic dynamics. The condition on jet behaviour allows us to
exclude objects such as the twin sources in \Ss{3C75} (Owen \etal\
1985; Hardcastle 1996) which are associated with a dominant cluster
galaxy and sometimes classed as WATs but whose inner jets are similar
to those of typical powerful FRIs.

Because of the requirements of this definition, wide-angle tail
sources make up a small minority of the radio source population. For
this reason, the detailed properties of their jets and tails have not
been well studied, although a number have been imaged for studies of
source dynamics (O'Donoghue \etal\ 1989). The only objects which have
been the subject of detailed study in the radio are \Ss{3C465} (Leahy
1984; Eilek \etal\ 1984) and \Ss{3C218}, Hydra A (Taylor \etal\ 1990),
although M87, Virgo A (e.g.\ Biretta \& Meisenheimer 1993) exhibits
some of the properties of a WAT. In this paper I present
multi-configuration, multi-frequency VLA observations of a further
powerful WAT.

Throughout this paper I use a cosmology in which $H_0 = 50{\rm\
km\,s^{-1}\,Mpc^{-1}}$ and $q_0 = 0$. At the distance of \Ss{3C130},
one arcsecond is equivalent to a projected length of
2.72 kpc. B1950.0 co-ordinates are used throughout.

\section{Observations}

\Ss{3C130} was observed with the VLA as part of a programme of detailed
observations of FRI radio galaxies. Dates and integration times are
shown in table \ref{obstab}. \Ss{3C286} and \Ss{3C48} were used as
primary flux calibrators; the nearby point sources 0537+531 and
0435+487 were used as phase calibrators, and (where \Ss{3C286} was not
observed) \Ss{3C138} was used as a polarization angle reference.
A bandwidth of 50 MHz was used, except at A configuration,
where 25 MHz was used to reduce bandwidth smearing.

The data were reduced within {\sc aips} in the standard way. The
datasets from each configuration were initially reduced separately,
each undergoing several iterations of CLEANing and phase
self-calibration.  The B, C and D-configuration datasets were then
phase-calibrated, using the appropriate baselines, with images made
from the higher-resolution datasets, with which they were then merged
without reweighting. Thus the B-configuration data were phase
calibrated with an image made from the A-configuration data and merged
with it to form an AB dataset; images made with this at low resolution
were used to phase calibrate the C-configuration data and the two were
merged to form an ABC dataset, and so on. This process ensures phase
consistency in the data while removing the need for a self-calibration
of the final merged dataset.

Maps were made using the {\sc aips} task IMAGR, with tapering of the
\uv\ plane where low-resolution maps were required. The robustness
parameter in IMAGR was used to temper the uniform weighting of the
\uv\ plane, to improve the signal-to-noise ratio. In all cases the
restoring beam was a circularly symmetrical Gaussian, well matched to
the Gaussian fit to the dirty beam, and the resolution quoted is its
FWHM. The total-intensity map at the highest resolution was made with
a combination of IMAGR and the maximum-entropy imaging task VTESS;
IMAGR was used to clean off the bright point-like components, the
residual image was deconvolved with VTESS, and the point-like
components subsequently restored.

\section{Results}

\subsection{Overall source structure}

\begin{figure*}
\begin{center}
\leavevmode
\epsfbox{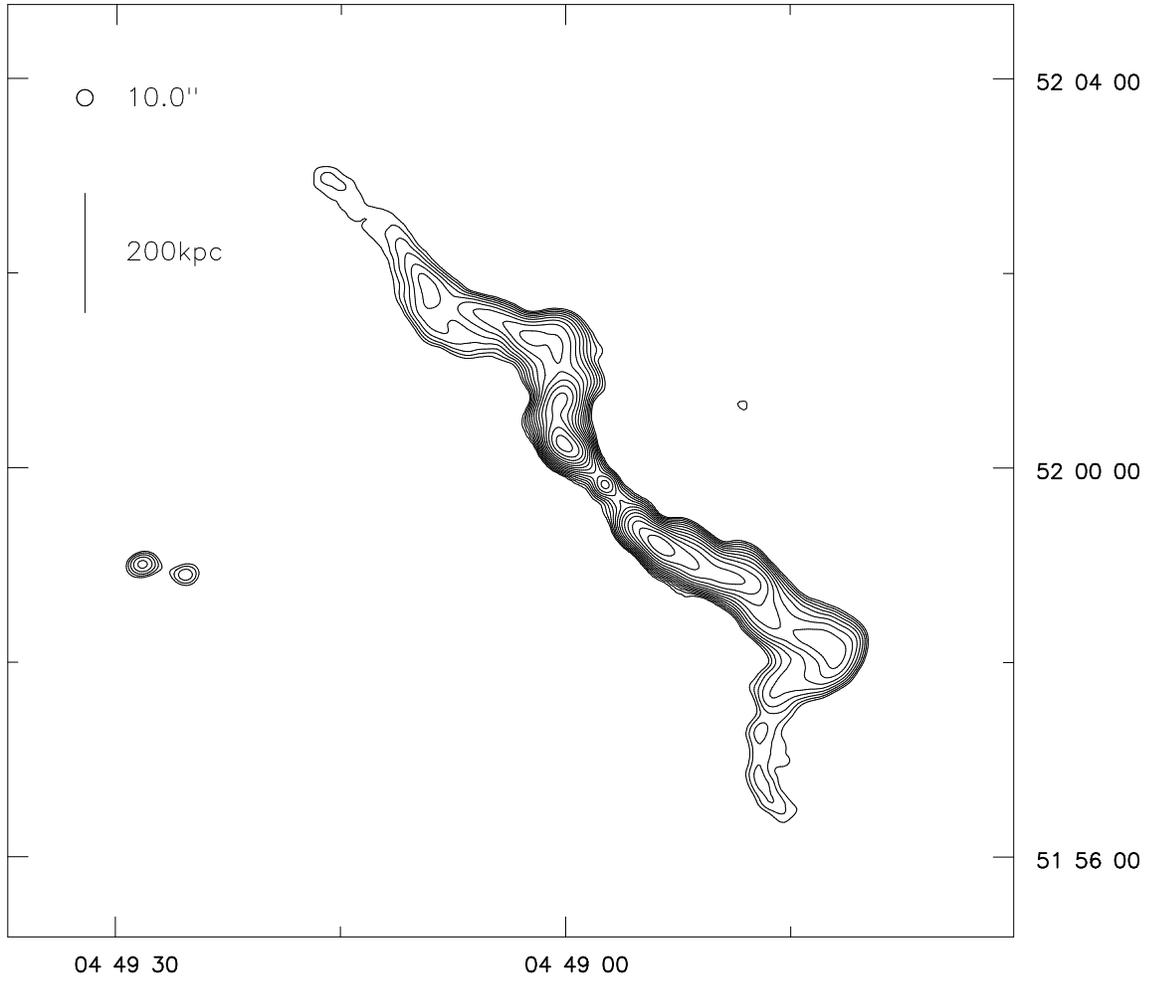}
\caption{1.5 GHz
map of \Ssf{3C130} at 10.0-arcsec resolution. Contours at $1.5 \times
(1, \protect\sqrt 2, 2, 2\protect\sqrt 2, \dots$) mJy
beam$^{-1}$.}
\label{3C130_low}
\end{center}
\end{figure*}

Fig.\ \ref{3C130_low} shows the large-scale structure of the
source. There are several pronounced bends, in spite of the overall
straightness of the source. The sudden change in direction at the end
of the south tail is particularly noticeable; this feature is similar to
several seen in the small sample of O'Donoghue, Owen and Eilek (1990).
The source disappears into the noise on these images and is longer
than the $\sim 1$ Mpc seen here.

\subsection{The core}
 
The radio core of \Ss{3C130} did not vary over the timescales of the
observations either at 8.4 or 1.5 GHz, within the errors imposed by
the uncertainty of absolute flux calibration at the VLA. Its flux at
8.4 GHz was 29.0 mJy and at 1.5 GHz 12.4 mJy.  The best position for the
core is RA 04 48 57.34, DEC +51 59 49.7. 

\subsection{The jets and hot spots}

\begin{figure*}
\begin{center}
\leavevmode
\vbox{\epsfysize 13cm\epsfbox{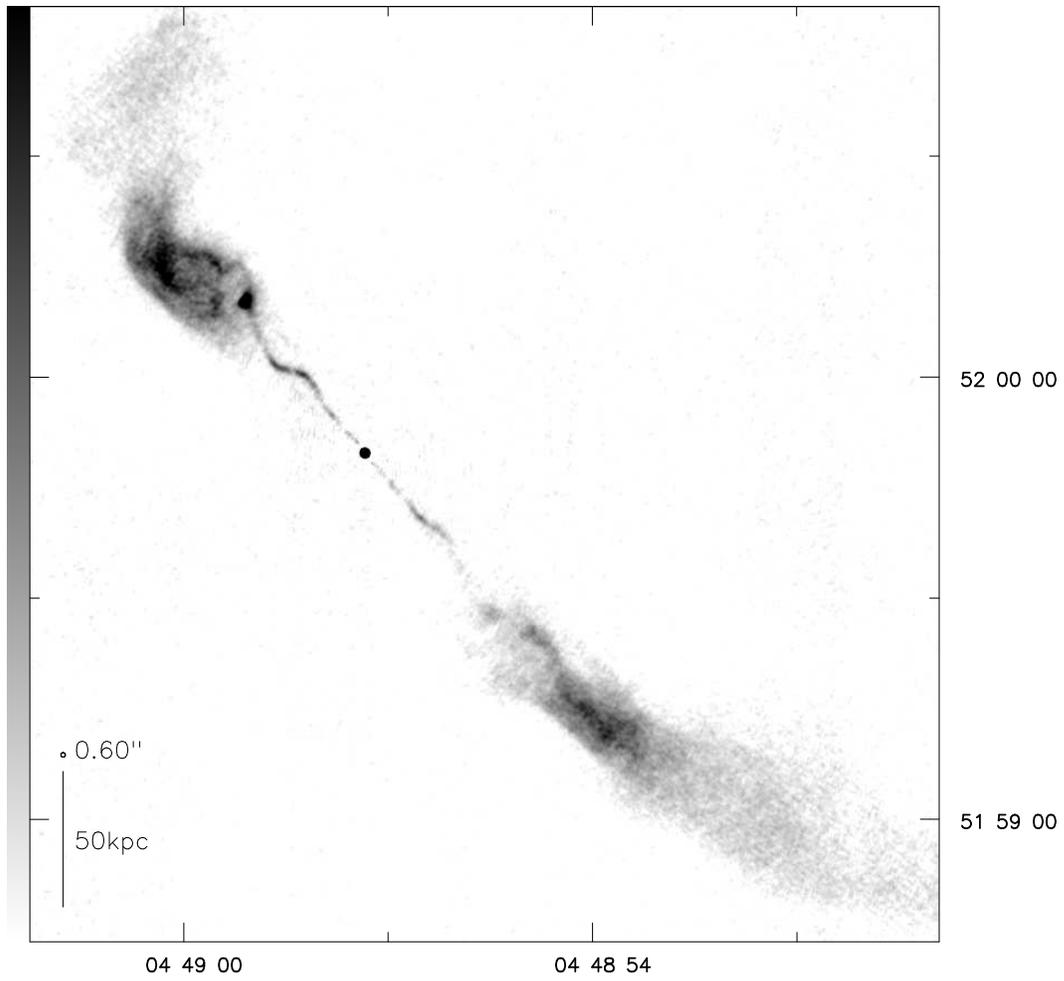}}
\caption{8.4-GHz map of \Ssf{3C130} at 0.60-arcsec resolution. Linear
greyscale; black is 0.4 mJy beam$^{-1}$.}
\label{3C130.060g}
\end{center}
\end{figure*}

\begin{figure*}
\begin{center}
\leavevmode
\vbox{\epsfysize 11.4cm\epsfbox{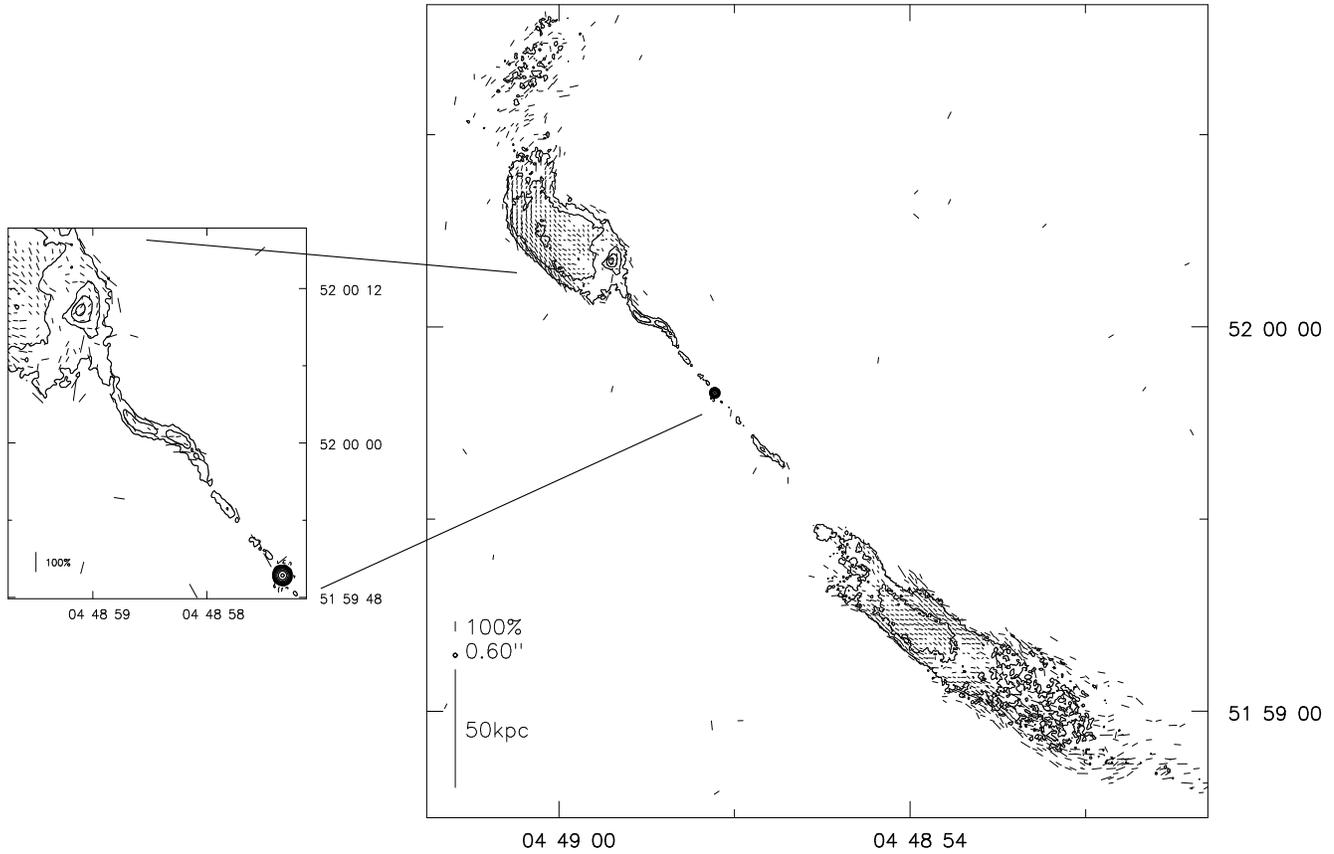}}
\caption{8.4-GHz map of \Ssf{3C130} at 0.60-arcsec
resolution. Contours at $0.1 \times (-2, -1,1, 2, 4, \dots)$ mJy
beam$^{-1}$. Vectors show inferred magnetic field direction and their
length is proportional to degree of polarization. The inset shows
details of polarization in the north jet and hot spot.}
\label{3C130.060c}
\end{center}
\end{figure*}

The high-resolution images in Figs \ref{3C130.060g} and
\ref{3C130.060c} show two very well-collimated jets emerging from the
core. The jets are reasonably symmetrical. The northern jet in
\Ss{3C130} is brighter, noticeably so at bends; over the inner section
where both jets are straight (approximately 9 arcsec) the difference
in brightness is roughly a factor 1.4. [This symmetry in the
brightness of jets is reasonably common among WAT sources, compared to
FRII radio galaxies or quasars (e.g.\ O'Donoghue \etal\ 1993). The
`archetype' of the class, \Ss{3C465}, appears to be
unusual in having a very one-sided jet.]  The bends in the jets,
particularly the northern one, are very striking. The beams may be
ballistic, implying some short-timescale wobble of the collimator
(`garden-hose' behaviour), but if this is the case it is surprising
that the jets are brighter at bends and that there is no antisymmetry
between the jet and counterjet. If they are not ballistic it is equally
remarkable that they remain collimated while undergoing oscillations
of such large amplitude in so short a distance. The northern jet
terminates in a hot spot, but there is a long filament which leaves
the hot spot to the north, possibly suggesting some continued
collimated outflow. At this resolution there is little compact
structure at the end of the southern jet; the `hot spot' seen in the
maps of Leahy (1985, 1993) is resolved, with a size of around a second
of arc. By contrast, the northern hot spot is only just resolved at
the full resolution of the dataset (0.24 arcsec; maps not shown) and
its brightest component has a minor axis of $\sim 0.3$ arcsec. This
use of the term `hot spot' is stronger than that of O'Donoghue \etal\
(1993), who only used it to indicate a brighter, broader region; the
hot spot seen here is comparable in compactness with those in nearby
FRIIs (e.g.\ Black \etal\ 1992; Leahy \etal\ 1997; Hardcastle \etal\
1997a) and is superposed on a brighter region which corresponds to the
`hot spot' of O'Donoghue et al. The northern jet is resolved at the
bends at full resolution, and has a cross-sectional width of up to 0.8
arcsec.

The polarization map (Fig.\ \ref{3C130.060c}) includes a correction
for Ricean bias and shows all points with polarized and total
intensity greater than three times the respective off-source r.m.s.\
noise values. The position-angle vectors are perpendicular to the
observed $E$-field, and so show the direction of the apparent magnetic
field if Faraday rotation is negligible. Although we expect a
non-negligible rotation measure (discussed further below, section
\ref{rotm}), these angles remain the best guess of the magnetic field
direction. On this basis, the jets have apparent magnetic field
parallel to their length where polarization is detected; the field
follows the bends in the northern jet. This is as expected for a
strong-flavour jet (e.g.\ Saikia \& Salter 1988). The field in the hot
spot is transverse to the jet direction and parallel to the hot spot's
direction of extension; this is similar to the field configuration in
many FRII hot spots (Hardcastle \etal\ 1997a) but also to that in the
termination knots of M87's jet (Owen, Hardee \& Cornwell
1989). Further out, the magnetic field is parallel to the plumes, and
the degree of polarization is high. This appears to be the behaviour
in the best-studied WATs (e.g.\ Taylor \etal\ 1990, O'Donoghue \etal\
1990, Patnaik \etal\ 1984; Saikia \& Salter 1988, and references
therein) but is quite different from the behaviour observed in the
weak-flavour jets of normal FRIs, in which the field is transverse to
the jet axis, sometimes with a longitudinal sheath (e.g.\ Hardcastle
\etal\ 1996; Laing 1996; Hardcastle \etal\ 1997b).

\subsection{Depolarization, rotation measure and spectral index}
\label{rotm}

Using matched-baseline maps, I confirm earlier findings that the
source is rapidly depolarized at low frequencies. The mean
depolarization between 1.5 and 8.4 GHz (averaged over the areas with
good signal-to-noise in both maps) of the northern plume is 0.2, and
that of the southern plume 0.1. It may be noteworthy that the southern
lobe, with a weaker jet and no bright compact hot spot, is the more
depolarized: this may be an example of a Laing-Garrington effect
(Laing 1988; Garrington \etal\ 1988) in WATs, although Saripalli
\etal\ (1996) suggest that there are substantial variations in the
degree of polarization with radio frequency. There is weak evidence
that the inner 50 arcsec of both lobes is more depolarized than the
outer parts, which would be consistent with depolarization by a medium
associated with the galaxy or cluster. There are no systematic
observations of depolarization in this class of source.

The rotation measure (RM) distribution is not constrained by the
rotation of polarization angle between 8.4 and 1.5 GHz. Rotations
through all possible angles take place over the source, so there are
variations in RM of more than 36 rad m$^{-2}$ on arcsecond
scales. This is consistent with the RM measurements of Leahy
(1985). Good maps at a higher frequency are needed to constrain the RM
distribution adequately. Saripalli \etal\ (1996) report measurements
suggesting an integrated galactic RM of $\sim 300$ rad m$^{-2}$ in the
region of \Ss{3C130}. From the fact that the polarization vectors are
well aligned with one another (and consistent with those in the
lower-resolution maps of Saripalli \etal ) in the 8.4-GHz maps, and
seem to follow bends in the source where these are present, we may
guess that the rotation measure towards any point in the source is not much
greater than this value, which would produce a $20\dgr$ rotation in
polarization position angle at 8.4 GHz.

\begin{figure*}
\begin{center}
\leavevmode
\vbox{\epsfysize 12cm\epsfbox{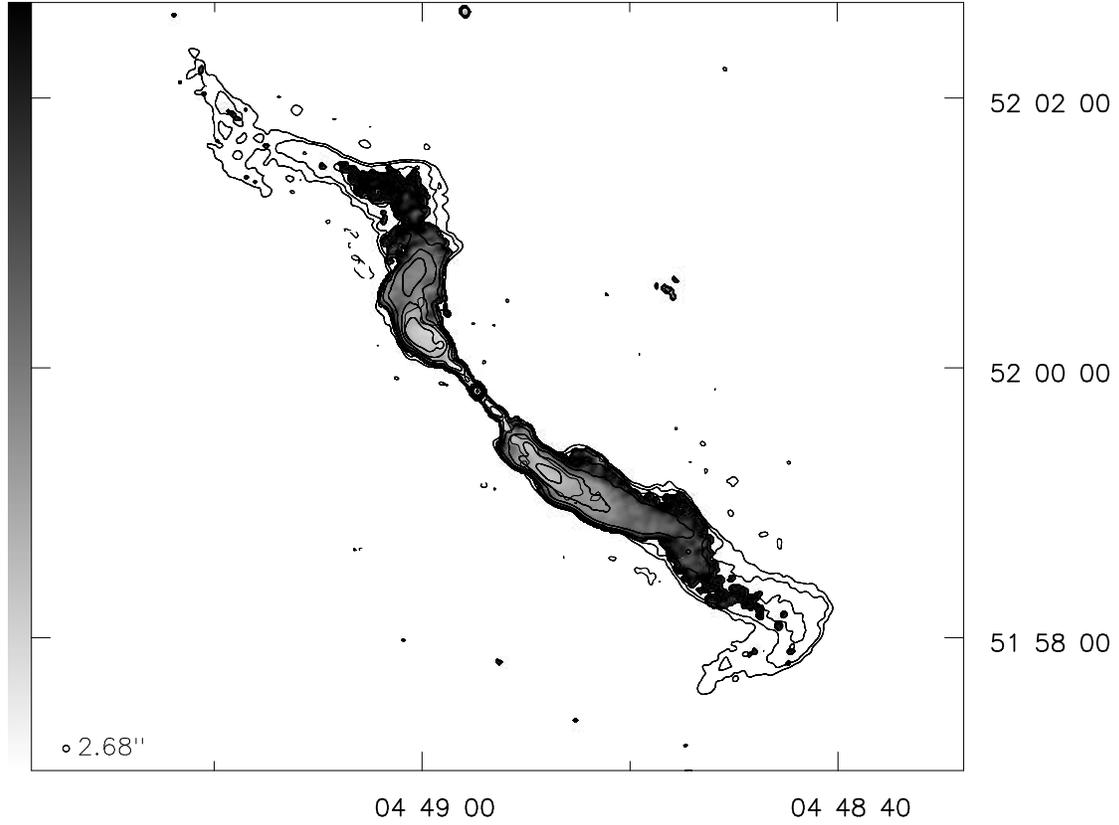}}
\caption{Spectral index between 8.4 and 1.5 GHz of
\Ssf{3C130} at 2.86-arcsec resolution. Linear greyscale between 0.4
and 1.5; superposed are contours of total intensity at 1.5 GHz at 
$0.4 \times (-2, -1, 1, 2, 
4, \dots)$ mJy beam$^{-1}$. }
\label{3C130-spix}
\end{center}
\end{figure*}

The spectral index of the source steepens rapidly with distance from
the core. Fig.\ \ref{3C130-spix} shows a map of spectral index; the
matched baselines of the maps ensure that the steepening is not an
effect of undersampling. This spectral behaviour is expected in the
standard model in which the plumes flow slowly away from the source
[compare the spectral index maps of Hydra A by Taylor \etal\
(1990)]. Note the comparatively flat ($\alpha \approx 0.5$) spectral
index of the jets and of the material they flow into. The northern hot
spot has a spectral index flatter than the material that surrounds it.

I determined spectral ages for regions along the (straighter) southern
tail, using a minimum energy for the relativistic electron
distribution corresponding to $\gamma = 100$, an initial
electron-energy power-law index of 2 to reflect the hot spot spectral
indices of 0.5, no energy contribution from relativistic protons,
filling factor unity, and equipartition magnetic fields; I took flux
measurements of regions of the plume between 30 and 110 arcsec,
measured along the plume, from the radio core. The ageing field used
was 0.46 nT, which was the mean of the equipartition fields fitted at
various points along the tail; there was little variation in
equipartition field strength with distance, so that this field is a
good approximation to the correct self-consistent value. The model
included the effects of inverse-Compton scattering from the CBR, which
at this redshift produces energy loss equivalent to that due to a
magnetic field of 0.40 nT. Using a model with effective pitch-angle
scattering of electrons (Jaffe \& Perola 1973), the plot of age
against projected distance along the source was well fitted by a
straight line with gradient $\sim 1.2 \times 10^4$ km s$^{-1}$,
inferred ages being of the order of $10^7$ years (as found by J\"agers
\& De Grijp 1985). The intercept was non-zero, reflecting the presence
of steeper-spectrum material surrounding the jet termination; derived
velocities were similar if the intercept was made zero by choosing a
steeper initial power-law index (2.53).  These inferred ages and
velocities in the tails are comparable to those found by spectral age
methods in other WAT sources (e.g.\ Taylor \etal\ 1990; O'Donoghue
\etal\ 1993) and would imply outflow which is considerably faster than
the sound speed in the external medium, given the temperature of the
gas around 3C\,130 (discussed below); this is perhaps surprising in
view of the absence of any evidence for post-hot spot shock structures
in the tails and of their generally relaxed appearance. The usual
caveats apply to velocities determined by spectral-ageing methods, but
it should be noted that most factors that can affect the velocity
(including a contribution to the energy density from relativistic
protons, a particle filling factor less than unity, and significant
projection of the radio source) will produce velocities higher than
the value given above. Only if the assumptions involved in the
spectral ageing analysis are seriously wrong -- for example, if there
is significant {\it in situ} particle acceleration in the tails or
significant magnetic field inhomogeneity -- can the tail velocity be
much lower than this value. Evidence for such processes is discussed
in Eilek (1996) and references therein.

\subsection{X-ray observations}

Miley \etal\ (1983) report on {\it Einstein}\/ IPC observations of
\Ss{3C130}. Serendipitously, the source is also included in the field
of a 39.4 ks {\it ROSAT} PSPC pointed observation, taken from the
public archives, of the X-ray emitting supernova remnant
RX\,04591+5147 (Pfeffermann, Aschenbach \& Predehl 1991; Reich \etal\
1992). Although the X-ray source associated with \Ss{3C130} is 32
arcmin away from the pointing centre of the PSPC, and is thus badly
vignetted, the observations have superior signal-to-noise to the {\it
Einstein}\/ data and show details of the X-ray structure of the
source. The cluster is detected at $2200 \pm 100$ PSPC counts between
0.1--2.4 keV (derived from a circle of 11 arcmin radius about the
centre of the X-ray source, using a background annulus between 11 and
17.5 arcmin), in spite of the reduced sensitivity of the PSPC at this
off-axis distance. Because of the difficulty of measuring the
background in the presence of extended emission from the SNR, and
because the shadows of the ring and one of the radial struts pass
close to the source, the derived count rate of $\sim 6 \times 10^{-2}$
s$^{-1}$ is uncertain. A rough correction for vignetting would imply
an on-axis count rate of $9 \times 10^{-2}$ s$^{-1}$. Using the
Post-Reduction Offline Software (PROS) within IRAF, I made spectral
fits to the data, correcting for the off-axis location of the
source. A single Raymond-Smith model provided a good fit ($\chi^2 =
15.8$ with 26 degrees of freedom), giving a best-fit temperature $kT =
2.9^{+9}_{-2}$ keV; the fitted galactic $N_H$ was $0.9^{+0.5}_{-0.2}
\times 10^{22}$ cm$^{-2}$ [cf. the value of $0.4 \times 10^{22}$ cm
$^{-2}$ predicted by interpolation from Stark \etal\ (1992)]. Errors
quoted for $N_H$ and $kT$ are $1\sigma$ for two interesting
parameters. Abundances were poorly constrained; 70 per cent solar
abundance gave marginally the best fit. With this best-fit model, the
0.1-2.4 keV luminosity of the cluster is $5 \times 10^{37}$ W,
consistent with the luminosity derived, on crude spectral assumptions,
from the {\it Einstein}\/ data by Miley \etal\ (1983); the cluster is
thus comparable in X-ray luminosity to rich Abell clusters, and the
temperature consistent with the temperature-luminosity relation
(e.g.\ David \etal\ 1993). There is no evidence for a lower temperature
in the central regions of the source, and so no evidence that a
cooling flow is present; this appears to be normal for the host
clusters of WATs (Norman, Burns \& Sulkanen 1988; G\'omez \etal\ 1997)
although Schindler \& Prieto (1997) suggest that a weak cooling flow
is present in Abell 2634, the host cluster of 3C\,465, and Hydra A
inhabits a cooling flow with high mass deposition rates (David \etal\
1990).

\begin{figure*}
\begin{center}
\leavevmode
\vbox{\epsfysize 10cm\epsfbox{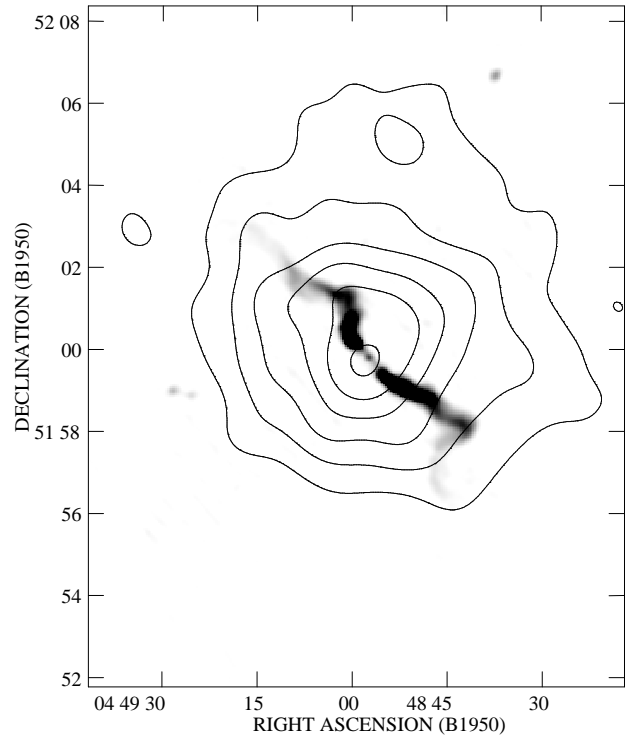}}
\caption{X-ray contours of the \Ss{3C130} cluster, smoothed with a
$\sigma = 70$-arcsec Gaussian and overlaid on a 10-arcsec resolution
greyscale of radio emission at 1.5 GHz (black is 25 mJy
beam$^{-1}$). The lowest contour is at background + $3\sigma$ and the
contour interval is $2\sigma$. No correction for differential
vignetting has been applied; the pointing centre of the PSPC
observation was 32 arcmin to the E.}
\label{3C130-xray}
\end{center}
\end{figure*}

The best-fit Gaussian to the off-axis point-spread function (PSF) of
ROSAT at this distance from the pointing centre has $\sigma \approx
70$ arcsec (Hasinger \etal\ 1995). For a radio-X-ray comparison I have
smoothed the broad-band (0.1-2.4 keV) X-ray image with a Gaussian of
this size; this should allow the coma-induced asymmetry of the PSF to
be neglected. The X-ray images (Fig.\ \ref{3C130-xray}) show an
extended structure on scales comparable to the length of the radio
source (i.e. $\sim 1$ Mpc). The cluster gas seems reasonably
symmetrical about the radio source, in contrast to the clumpy
structures, with offset radio sources, seen in some lower-luminosity
WAT hosts even at lower spatial resolution (e.g.\ Burns \etal\ 1994;
G\'omez \etal\ 1997). The distortion of the X-ray isophotes to the
northeast coincides with, and may be related to, the kink ($\sim
150$--300 kpc from the nucleus) in the northern tail; there is no
structure in the X-ray emission which can be related to the sudden
change in direction at the end of the southern radio tail, however.
Given the large and asymmetrical PSF, I have not attempted to fit
radial profiles to the X-ray data.

\section{Discussion}

Approaches to the source dynamics of WATs in the literature (e.g.\
Burns 1981; Eilek \etal\ 1984; O'Donoghue \etal\ 1993) have
concentrated on the large-scale bends seen in the tails. It is
instructive to ask a rather different question: why are these sources,
with well-collimated strong-flavour jets, compact hot spots, and high
radio luminosities, not classical double radio galaxies? In FRII
objects of this radio power, radio-emitting plasma is thought to flow
back from the hot spots into the `cocoon' left behind as the hot spot
and associated shocks propagate into the external medium, forming the
radio lobes (e.g.\ Scheuer 1974; Williams 1991). In WATs, the jet
appears to terminate in a shock in the same way. Norman \etal\ (1988)
argue that strong shocks are necessary to explain the single-step
transition between jets and plumes, and observations of compact hot
spots in these objects, such as that seen in 3C\,130, support this
model. However, the situation after the shock is different in the two
classes of object. In WATs lobes are not formed. Instead, the hot spot
is at the base of the tail; by analogy with the standard model for
FRIIs, we may assume that the emitting material in the tail has passed
through and been excited in the hot spot, and this is borne out by the
spectral index results in 3C\,130. The tails may immediately
deviate from the axis defined by the jets (e.g.\ \Ss{3C465}) or appear
to continue in a straight line (e.g.\ \Ss{3C130}) but in no case does
there appear to be lobe emission {\it significantly} closer to the
core than the hot spot.\footnote{Whether the bends in the jet seen in
3C\,130 are due to ballistic motion or to buffeting by the IGM, it is
clear that the position of the jet termination point, however it is
formed, must change with time. The hot spot will therefore move about
in the base of the plume in a manner similar to that described in the
`dentist's drill' model of Scheuer (1982) for the end points in
FRIIs. We do not therefore expect to see the hot spot at a particular
place in the tail in all cases.} The fact that there is no cocoon may
explain the brightness and two-sidedness of the strong-flavour jets in
WATs compared to those in FRIIs; a direct interaction with the
(comparatively dense) external medium might be expected to make the
beam more dissipative and perhaps to slow the regions of the beam
responsible for the emission to only weakly relativistic
velocities. This would explain the low values ($0.2c$) of `jet
velocity' estimated from sidedness by O'Donoghue \etal\ (1993)
compared to the much higher values (0.6--$0.7c$) estimated from the
sidedness of jets in FRII quasars (Bridle \etal\ 1994; Wardle \& Aaron
1997) and their prominence and sidedness in FRII radio galaxies
(Hardcastle \etal\, in prep.). Hardcastle \etal\ (1997a) proposed a
similar explanation for the prominence and two-sidedness of the jets
in the peculiar FRII 3C\,438. However, in the absence of classical
double lobes and the associated discontinuity between radio-emitting
plasma and shocked external medium, why are there jet termination
shocks in WATs?

It is well known that the difference between WATs and classical doubles is
the local environment; whereas WATs always lie at the centres of
clusters, FRII radio galaxies of comparable powers tend to avoid them
(e.g.\ Prestage \& Peacock 1988). An explanation for the peculiar
properties of WATs compared to their classical double counterparts
must turn on this environmental difference. A suggestion along these
lines by Leahy (1984), applied to \Ss{3C465}, invoked motion of the
galaxy through the cluster, causing it to leave behind a passive wake
of radio-emitting material; in this type of model the post-hot-spot
material is left behind by the motion of the galaxy and so never forms
a lobe. However, the motions of central cluster galaxies are not
expected to be large (Eilek 1984; Pinkney \etal\ 1993 and references
therein) and in any case such a model cannot account, without invoking
projection effects implausibly often, for the large population of WATs
in which one or both tails are more or less aligned with the inner
jets (as in 3C\,130).

Burns \etal\ (1994) suggest a model in which WATs have an origin in
the merger of a cluster with a group or subcluster. This is motivated
by the X-ray substructure which they find in many WAT host
clusters. Large-scale, high-velocity residual motions of gas could
then be responsible for the bending of the radio tails, while the
merger would provide tidally stripped gas to fuel the AGN. In an
extension of this work G\'omez \etal\ (1997) show that the majority of
WAT hosts in a larger sample show some X-ray substructure, with an
alignment between the direction of the X-ray elongation and the angle
that bisects the tails, consistent with such a model. 3C\,130,
however, is clearly a WAT despite the location of its host at the
centre of a smooth, approximately symmetrical distribution of X-ray
emitting gas and its (apparently) straight tails. It appears that
strong cluster inhomogeneity, though it may be necessary for bent tail
formation, is not necessary for the existence of a WAT; in particular
it does not, on its own, explain the jet shock/hot spot behaviour
discussed above.

Loken \etal\ (1995) discuss the physics of a jet propagating across
the boundary between the interstellar medium of the host galaxy and
the hotter, less dense intracluster medium, and suggest that this may
be the reason for the disruption of the inner, well-collimated jet at
a hot spot. They then postulate velocity shear across the boundary, as
described above, to account for jet bending. The structures seen in
numerical simulation when the jet simply crosses a contact
discontinuity with crosswind do not resemble WATs strongly,
however. If the jets are taken to cross a shock front instead (cf.\
Norman \etal\ 1988), then the simulations of Loken \etal\ are more
convincing in their resemblance to WATs, but we again face the problem
of the smoothness of the large-scale X-ray emission in 3C\,130; there
is little evidence in this source for the recent cluster merger that
Loken \etal\ invoke to produce such a shock. Neither the sonic radius
of a possible cooling flow nor the shock front associated with a
putative nuclear or galactic wind are in the appropriate place to
produce the internal shocks in WAT jets (Soker \& Sarazin 1988; Smith,
Kennel \& Coroniti 1993). Because of the low resolution of the X-ray
data, cluster-merger models cannot be ruled out for 3C\,130. Producing
such straight plumes in such a model while still causing both jets to
disrupt requires a rather special geometry for the merger and/or
convenient projection effects, however.

If hot spots in WATs represent jet termination shocks, it is perhaps
surprising that only a single hot spot is seen in 3C\,130 and that
there are no clear hot spot candidates in several of the sources of
O'Donoghue \etal\ (1993). It is possible that there are intrinsically
similar hot spots but that relativistic beaming effects are affecting
their visibility. In 3C\,130 the hot spot in the N lobe is
approximately ten times brighter than the most comparable feature in
the S lobe, which using standard results requires flow or advance
velocities greater than $0.3c$. More high-resolution observations of
these objects are needed to test such a model.

\section{Conclusions}

A compact hot spot is detected at the base of one plume of the WAT
\Ss{3C130}, and the jets are shown to have longitudinal magnetic
field. The source is thus very like a classical double in some
respects. The data support the model in which WATs are objects whose
jets make the transition from super- to sub-sonic velocities in one
step, rather than decelerating gradually, by showing a bright sub-kpc
structure (comparable to those seen in classical double radio sources)
associated with the termination of a jet.

Archival {\it ROSAT} PSPC observations of 3C\,130 show it to lie in a
luminous cluster with $kT \sim 2.9$ keV. There is little sign of
substructure in the X-ray, in contrast to many other WATs; this may be
related to the nearly straight tails of 3C\,130. The lack of strong
substructure seems to be inconsistent with recent models
for jet disruption in WATs.

\section*{ACKNOWLEDGEMENTS}
 
I am grateful to Julia Riley and Guy Pooley for suggesting the
original radio observations of this source, and thank Mark Birkinshaw,
Julia Riley and Diana Worrall for useful comments. I acknowledge a
research studentship from the UK Particle Physics and Astronomy
Research Council (PPARC) and support from PPARC grant GR/K98582. The
National Radio Astronomy Observatory is operated by Associated
Universities Inc., under co-operative agreement with the National
Science Foundation. This project made use of Starlink facilities. This
research has made use of the NASA/IPAC Extragalactic Database (NED)
which is operated by the Jet Propulsion Laboratory, California
Institute of Technology, under contract with the National Aeronautics
and Space Administration. This research has made use of data obtained
from the High Energy Astrophysics Science Archive Research Center
(HEASARC), provided by NASA's Goddard Space Flight Center.

\bsp
\clearpage
\begin{table}
\caption{VLA observations of \Ss{3C130}}
\label{obstab}
\begin{center}
\begin{tabular}{llrlr}
&\multicolumn{2}{c}{8.4 GHz}&\multicolumn{2}{c}{1.5 GHz}\\
Conf.&Date&$t_{int}$&Date&$t_{int}$\\
&&(mins)&&(mins)\\
A&1995/08/06$^a$& 120&1995/07/23$^a$&45\\
B&1995/11/28&120&1995/11/28&30\\
C&1994/11/10&55&1994/11/10&50\\
D&1995/03/06&20&1995/03/06&15\\
\end{tabular}
\end{center}
\parbox{\linewidth}{\medskip $^a$ Bandwidth of 25 MHz used.\par
$t_{int}$ denotes the total time spent on source at the specified VLA
configuration and frequency.}
\end{table}

\clearpage

\end{document}